\documentclass[12pt]{article}
\usepackage{epsfig}

\newcommand{\be}{\begin{equation}}
\newcommand{\bea}{\begin{eqnarray}}
\newcommand{\en}{\end{equation}}
\newcommand{\eea}{\end{eqnarray}}

\newcommand{\fract}[2]{{\textstyle\frac{#1}{#2}}} 

\def\thebibliography#1{\centerline{\large \bf References}\list
  {[\arabic{enumi}]}{\settowidth\labelwidth{[#1]}\leftmargin\labelwidth
    \advance\leftmargin\labelsep
    \usecounter{enumi}}
    \def\newblock{\hskip .11em plus .33em minus .07em}
    \sloppy\clubpenalty4000\widowpenalty4000}

\begin{document}

\begin{center}
{\LARGE\bf Soliton Picture for Pentaquarks\footnote{Talk presented 
at the mini workshop, {\it EXCITING HADRONS}, Bled, July, 2005}
}

\bigskip

{\large Herbert Weigel}

\smallskip
{Fachbereich Physik, Siegen University\\
Walter--Flex--Stra{\ss}e 3, D--57068 Siegen, Germany}

\bigskip

{\bf Abstract}

\smallskip

\parbox[t]{15.3cm}{\small
In this talk I report on a thorough comparison between the
bound state and rigid rotator approaches to generate baryon 
states with non--zero strangeness in chiral soliton models.
This comparison shows that the scattering amplitude in the 
bound state approach contains contributions generated by 
the exchange of the rigid pentaquark excitation, and that the 
two approaches are consistent with each other in the large $N_C$ limit.
The comparison paves the way to unambiguously compute the 
width of the~$\Theta^+$ pentaquark in chiral soliton models.}

\end{center}

\bigskip
\centerline{\large \bf Introduction}
\smallskip

In this talk I have discussed two issues regarding pentaquarks in chiral 
soliton models. First I have reviewed the relation between exotic 
five--quark states and radial excitations. In particular I have 
explained that the wave--functions of the crypto--exotic partners 
of the pentaquarks have significant admixture of radial excitations 
of the ordinary baryons and that this may have significant impact 
on transition magnetic moments. I have extensively described that 
issue before~\cite{We98} and will abstain from repeating it in 
these proceedings. Rather, I will focus on the second topic of my 
talk which deals with potential differences between the bound state 
and rigid rotator approaches (BSA and RRA, respectively) to generate 
baryon states with non--zero strangeness from the classical soliton; 
two seemingly different treatments of the \emph{same} model. It has 
previously been argued that the prediction of pentaquarks, {\it i.e.\@} 
exotic baryons with strangeness $S=+1$, would be a mere artifact of 
the RRA~\cite{It04}. A major result of the investigation presented in 
this talk is that pentaquark states do indeed emerge in both approaches. 
This comparison furthermore shows how to unambiguously compute the width 
of pentaquarks. That computation of the width differs substantially 
from previous approaches based on assuming pertinent transition operators 
for $\Theta^+\to KN$~\cite{Di97,El04}. Details of these studies and an 
exhaustive list of relevant references are contained in the recent 
paper~\cite{Wa05} in collaboration with Hans Walliser.

The qualitative results, on which I focus, 
are \emph{model independent} while quantitative results may be quite 
sensitive to the model parameters and/or the actual form of the chiral 
Lagrangian. For simplicity, our calculations in ref.~\cite{Wa05}
have been performed in the Skyrme model augmented by the Wess--Zumino 
and the simplest flavor symmetry breaking terms. The latter parameterizes 
the kaon--pion mass difference. 

Chiral soliton calculations are organized in powers of $N_C$, the number 
of colors and hidden expansion parameter of QCD. The leading contribution
is the classical soliton energy, $E_{\rm cl}={\cal O}(N_C)$. 
The reported calculation is complete to ${\cal O}(N_C^0)$,
identifies the resonance contribution in kaon--nucleon
scattering and provides insight in $1/N_C$ corrections.

\bigskip
\centerline{\large \bf Small amplitude vs. collective coordinate quantization}
\smallskip

I start with phrasing the problem and briefly
review the two popular approaches to generate baryon states with 
strangeness $S=\pm1$ from a soliton configuration.
Chiral soliton models are in general functionals of the chiral field,~$U$,
the non--linear realization of the pseudoscalar fields.
Starting point in these considerations is the classical
soliton, {\it i.e.\@} the hedgehog embedded in the isospin
subgroup of flavor $SU(3)$,
\begin{equation}
U_0(\vec{x\,})={\rm exp}\left[i\vec{\tau\,}\cdot\hat{x} F(r)\right]\,,
\quad r=|\vec{x\,}|
\label{hedgehog}
\end{equation}
parameterized by the three Pauli matrices $\tau_i$.
The essential issue, however, is the treatment of the strange degrees of
freedom, the kaons.

The \emph{ansatz} for small amplitude quantization of kaon modes, known as 
BSA, reads
\begin{equation}
U(\vec{x\,},t)=A_2(t) \sqrt{U_0(\vec{x\,})}\,\,
{\rm exp}\left[\frac{i}{f_\pi}\sum_{\alpha=4}^7\lambda_\alpha
\eta_\alpha(\vec{x\,},t)\right]
\sqrt{U_0(\vec{x\,})}\, A_2^\dagger(t)\,,
\label{Usu2}
\end{equation}
where $\lambda_\alpha$ are Gell--Mann matrices of $SU(3)$. The small 
amplitude fluctuations $\eta_\alpha$ are treated in harmonic 
approximation. The pion decay constant, $f_\pi$ is ${\cal O}(\sqrt{N_C})$. 
Hence this harmonic expansion is complete at ${\cal O}(N_C^0)$. Subleading 
contributions may be substantial but they are not under control in the BSA. 
The dynamical treatment of the collective coordinates, $A_2\in SU(2)$ for 
the spin--isospin orientation of the soliton adds some of them.
Quantization of both $\eta_\alpha$ and $A_2$ results in the mass formula
\begin{equation}
M_S = E_{\rm cl}+ \omega_S+\frac{1}{2\Theta_\pi}\left[
c_SJ\left(J+1\right)+\left(1-c_S\right)I\left(I+1\right)\right]
+{\cal O}\left(\eta^4\right)\,.
\label{MBSA}
\end{equation}
for strangeness $S=\pm1$ baryons.
Here $J$ and $I$ are the spin and isospin quantum numbers of the 
considered baryon, respectively. The parameters in eq.~(\ref{MBSA})
can be approximated as functionals of the chiral angle, $F(r)$ 
and are conveniently described by defining 
$\omega_0=\frac{N_C}{4\Theta_K}$,
\begin{equation}
\omega_{\pm}=\frac{1}{2}\left[\sqrt{\omega_0^2+\frac{3\Gamma}{2\Theta_K}}
\pm\omega_0\right]
\quad {\rm and} \quad
c_{\pm}=1-\frac{4\Theta_\pi\omega_\pm}{8\Theta_K\omega_\pm \mp N_C}\,.
\label{cpara}
\end{equation}
The difference between $\omega_+$ and $\omega_-$ originates from
the Wess--Zumino term.
Explicit expressions for the moments of inertia $\Theta_\pi$ (rotation
in coordinate space) and $\Theta_K$ (rotations in $SU(3)$ flavor space)
as well as the symmetry breaking parameter $\Gamma$ (proportional
to $m_K^2-m_\pi^2$) may be traced from the literature~\cite{Wa05}.
They are all ${\cal O}(N_C)$.

\begin{figure}[t]
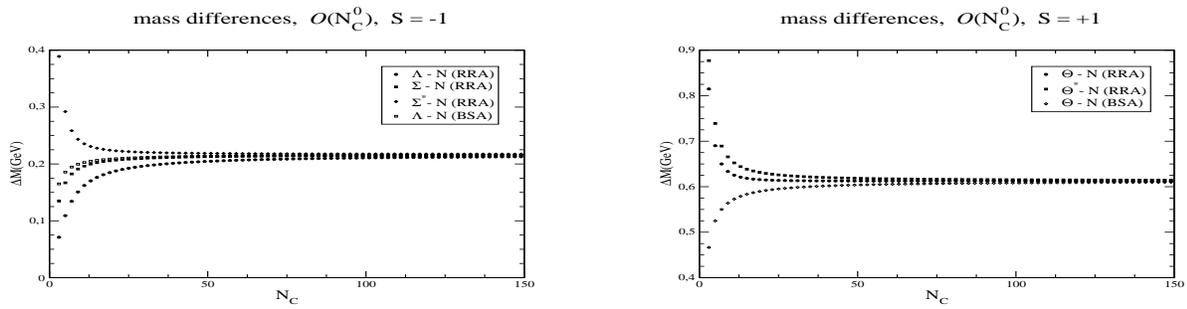

\centerline{\hskip -0.0cm
\epsfig{figure=mass_0m.eps,height=4.0cm,width=7.0cm}\hspace{1.5cm}
\epsfig{figure=mass_0p.eps,height=4.0cm,width=7.0cm}}
\caption{\label{fig_1}\sl \footnotesize
Mass differences at ${\cal O}(N_C^0)$ computed within the bound state 
and rigid rotator approaches (BSA and RRA, respectively)
in the Skyrme model as functions of $N_C$. In the
RRA they are the corresponding differences of the eigenvalues
in eq.~(\ref{evprob}) while in the BSA they are extracted from
eq.~(\ref{MBSA}). Left panel $\Delta S=-1$; right panel
$\Delta S=+1$.}
\end{figure}

\begin{figure}[t]
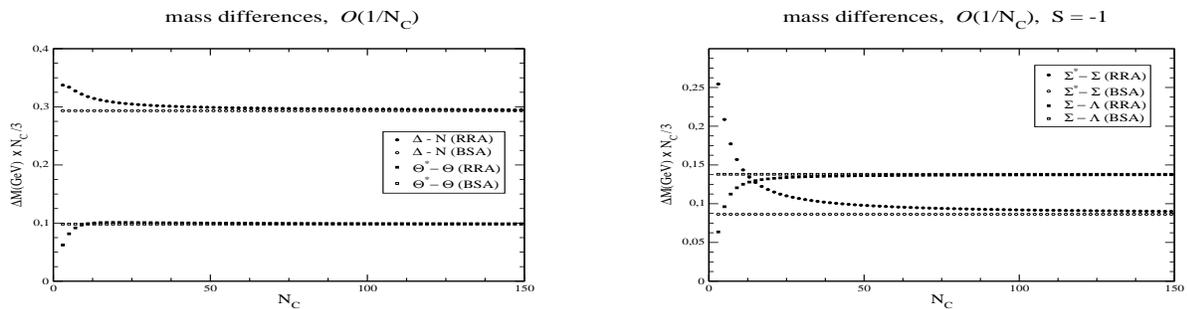

\centerline{\hskip -0.0cm
\epsfig{figure=mass_1p.eps,height=4.0cm,width=7.0cm}\hspace{1.5cm}
\epsfig{figure=mass_1m.eps,height=4.0cm,width=7.0cm}}
\caption{\label{fig_2}\sl \footnotesize
Mass differences at ${\cal O}(1/N_C)$ computed 
in the Skyrme model as functions of $N_C$.
Left panel: baryons with strangeness $S=0,1$; right panel $S=-1$.
See also caption of fig.~\ref{fig_1}.}
\end{figure}

The second approach treats the kaon modes purely as collective 
excitations of the classical soliton, eq.~(\ref{hedgehog}). These
collective modes are maintained to all orders and quantized 
canonically. The {\it ansatz} for this so--called rigid rotator 
approach (RRA) reads
\begin{equation}
U(\vec{x\,},t)=A_3(t) U_0(\vec{x\,})\, A_3^\dagger(t)
\quad {\rm with} \quad A_3(t)\in SU(3)\,.
\label{Urra}
\end{equation}
This parameterization describes only a limited number
of soliton excitations, those that arise as a rigid rotation of the
classical soliton. Though generating contributions of ${\cal O}(N_C^0)$
to baryon masses it is not complete at this order, {\it e.g.\@}
S--wave excitations are not accessible. However, since the rigid 
rotations are treated to any order, they control signifi\-cant 
subleading effects on the low--lying P--wave baryons. 
{}From the {\it ansatz}, eq.~(\ref{Urra}) the Hamil\-tonian 
for the collective coordinates is straightforwardly derived. The 
corresponding baryon spectrum is the solution to the 
eigenvalue problem ($D_{ab}=\frac{1}{2}{\rm tr}
[A \lambda_a A^\dagger \lambda_b]$)
\begin{equation}
\left\{\left(\frac{1}{2\Theta_\pi}-\frac{1}{2\Theta_K}\right)J(J+1)
+\sum_{a=1}^7\frac{R_a^2}{2\Theta_K}
+\frac{\Gamma}{2}\left(1-D_{88}\right)\right\}\Psi
={\cal E}\Psi\,,
\quad 
R_8\Psi=\frac{N_C}{2\sqrt3}\Psi\,,
\label{evprob}
\end{equation}
in each spin--isospin channel. The $R_a$ denote the (intrinsic) 
$SU(3)$ generators 
conjugate to the collective rotations $A_3\in SU(3)$. This eigenvalue 
problem is (numerically) exactly solved for arbitrary (odd) $N_C$ and 
symmetry breaking 
$\Gamma$ by generalizing the techniques of ref.~\cite{Ya88}. Then the 
eigenvalues ${\cal E}$ determine the baryon spectrum. In the flavor 
symmetric case the eigenstates are members of $SU(3)$ representations. 
For $N_C=3$ those are the octet and the decuplet for the low--lying 
$J=\fract{1}{2}$ and $J=\fract{3}{2}$ baryons, respectively. Also 
states in the anti--decuplet, $\mathbf{\overline{10}}$ are low--lying.
Probably the lowest 
mass state in the $\mathbf{\overline{10}}$ is the $\Theta^+$ pentaquark.
For arbitrary $N_C$ the condition on $R_8$ alters the allowed
$SU(3)$ representations and the inclusion of flavor symmetry
breaking leads to mixing of states from different representations.
These effects are incorporated in the exact numerical solution.
In figs.~\ref{fig_1} and \ref{fig_2} I compare the spectra for the 
low--lying P--wave baryons obtained from eqs.~(\ref{MBSA}) 
and~(\ref{evprob}) as functions of $N_C$.
Obviously the two approaches yield identical results as 
$N_C\to\infty$, as they should. This is the case for the ordinary hyperons
and the pentaquarks. Fig.~\ref{fig_2} also shows 
that even with flavor symmetry breaking included, the $\Delta$--nucleon
mass difference is ${\cal O}(1/N_C)$ in contrast to what is stated in
ref.~\cite{Pr03}.

\bigskip
\centerline{\large \bf Constrained fluctuations and $\mathbf{\Theta^+}$
width}
\smallskip

The above observed identity between BSA and RRA in the large $N_C$ limit 
has a caveat. Though $\omega_{-}<m_K$ corresponds to a true bound state,
$\omega_+$ is a continuum state. Thus, a pronounced resonance 
structure is expected in the corresponding phase shift. However,
that is not the case, as indicated in the left panel of
fig.~\ref{fig_3}. The computed phase shift hardly reaches 
$\pi/2$ rather then quickly passing through this value.
This has been used to argue that pentaquarks are
a mere artifact of the RRA~\cite{It04}. However, the ultimate 
comparison requires to generalize the RRA to the 
rotation--vibration approach (RVA)
\begin{equation}
U(\vec{x\,},t)=A_3(t) \sqrt{U_0(\vec{x\,})}\,
{\rm exp}\left[\frac{i}{f_\pi}\sum_{\alpha=4}^7\lambda_\alpha
\widetilde{\eta}_\alpha(\vec{x\,},t)\right]
\sqrt{U_0(\vec{x\,})}\, A_3(t)^\dagger\,.
\label{Usu3}
\end{equation}
Modes that correspond to the collective rotations must be 
excluded from the fluctuations $\widetilde{\eta}$, {\it i.e.\@}
the fluctuations must be orthogonal to the zero--mode
$z(r)\sim {\rm sin}\left(\frac{F(r)}{2}\right)$.  Imposing
the corresponding constraints for these fluctuations (and
their conjugate momenta) yields integro--differential equations
listed in ref.~\cite{Wa05}. For the moment let's omit the coupling
between $\widetilde{\eta}$ and the collective soliton excitations
(eigenstates of eq.~(\ref{evprob}), {\emph{including} pentaquarks). This 
truncation defines the background wave--function $\overline{\eta}$ 
(also orthogonal to the zero mode). Treating $\overline{\eta}$ 
as an harmonic fluctuation provides the  background phase 
shift shown as the blue curve in the right panel 
of fig.~\ref{fig_3}. Remarkably, the difference between the 
phase shifts of $\overline{\eta}$ and $\eta$ exhibits a clear
resonance structure. It is the resonance phase shift to be associated 
with the $\Theta^+$ pentaquark in the limit $N_C\to\infty$.

There is an even more convincing computation of this resonance
phase shift. In contrast to the parameterization, eq.~(\ref{Usu2})
the {\it ansatz}, eq.~(\ref{Usu3}) yields an interaction Hamiltonian
that is linear in the fluctuations, generating Yukawa couplings between 
the collective soliton excitations and the fluctuations $\widetilde{\eta}$. 
In ref.~\cite{Wa05} we have derived this Hamiltonian keeping all 
contributions that survive as $N_C\to\infty$. The corresponding Yukawa 
exchanges extend the integro--differential equations for $\overline{\eta}$ 
by a separable potential $V_Y$, therewith providing the equations of 
motion for $\widetilde{\eta}$~\cite{Wa05}. The equation of motion for 
$\widetilde{\eta}$ is solved by $\widetilde{\eta}=\eta-az$ with
$a=\langle z|\eta\rangle$ for $N_C\to\infty$~\cite{Wa05},
where $\eta$ is the unconstrained small
amplitude fluctuation of the BSA, eq.~(\ref{Usu2}). The phase shifts
extracted from $\eta$ and $\widetilde{\eta}$ are identical because
$z(r)$ is localized in space.
\begin{figure}[t]
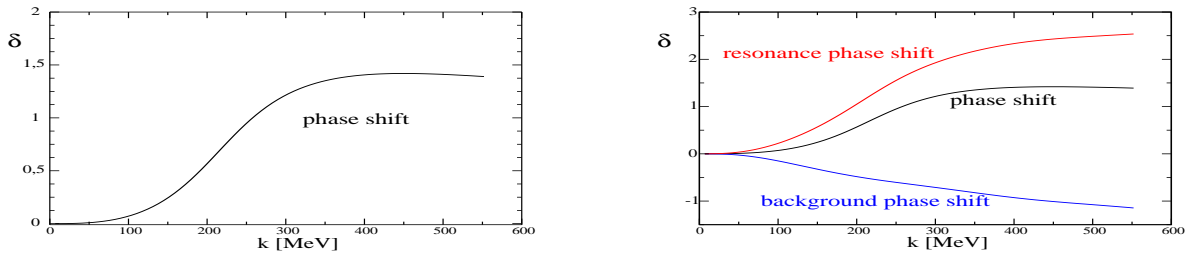

\centerline{
\epsfig{file=phaseshift.eps,height=4cm,width=7cm}\hspace{1.5cm}
\epsfig{file=resonance.eps,height=4cm,width=7cm}}
\caption{\label{fig_3}\sl \footnotesize
Phase shift computed in the BSA (left panel)
and resonance phase shift after removal of the background
contribution in the RVA (right panel). Note the different scales.}
\end{figure}
Thus the BSA and RVA yield the same spectrum and are indeed equivalent 
in the large $N_C$ limit. But, the RVA provides a distinction between 
resonance and background contributions to the scattering amplitude.
Applying the $R$--matrix formalism on top
of the constrained fluctuations $\overline{\eta}$ shows that $V_Y$
\emph{exactly} contributes the resonance phase shift shown
in fig.~\ref{fig_3} when the Yukawa coupling is computed for
$N_C\to\infty$. This identifies the exchange of a state predicted
in the RRA which thus is no artifact. In contrast, pentaquarks
are also predicted by the BSA; just well hidden. However, collective 
coordinates are mandatory to obtain finite $N_C$ corrections to the BSA
for the properties of~$\Theta^+$. Though not all ${\cal O}(1/N_C)$
operators were included in ref.~\cite{Wa05}, subleading effects have
turned out to be substantial. For example, in the case $m_K=m_\pi$ the 
mass difference with respect to the nucleon increases by a factor two
from $\omega_0$ to $(N_C+3)/4\Theta_K$ for $N_C=3$. In the 
realistic case with $m_K\ne m_\pi$ this mass difference is obtained from 
solving the eigenvalue problem, eq.~(\ref{evprob}). Furthermore, the 
resonance (extracted from the comparison between $\widetilde{\eta}$
and $\overline{\eta}$) becomes sharper as $N_C<\infty$~\cite{Wa05}.

The separable potential $V_Y$ also provides the general expression 
for the width as a function of the kaon energy
$\omega_k=\sqrt{k^2+m_K^2}$ from the $R$--matrix formalism~\cite{Wa05}
\begin{equation}
\Gamma(\omega_k)=2k\omega_0
\left|X_\Theta\int_0^\infty r^2dr\, z(r)2\lambda(r)
\overline{\eta}_{\omega_k}(r)
+\frac{Y_\Theta}{\omega_0}\left(m_K^2-m_\pi^2\right)
\int_0^\infty r^2dr\,z(r)\overline{\eta}_{\omega_k}(r)\right|^2\,.
\label{widthsb}
\end{equation}
Here $\overline{\eta}_{\omega_k}(r)$ is the P--wave projection 
of the background wave--function $\overline{\eta}$ for a prescribed
energy $\omega_k$. Furthermore $\lambda(r)$ is a 
radial function that stems from the Wess--Zumino term. 
The matrix elements of the collective coordinate operators 
that enter in eq.~(\ref{widthsb}) ($D_{\pm a}=D_{4a}\pm iD_{5a}$)
\begin{equation}
X_\Theta:=\sqrt{\frac{32}{N_C}}
\langle \Theta^+| \sum_{\alpha,\beta=4}^7d_{3\alpha\beta}
D_{+\alpha}R_\beta|n\rangle 
\,,\qquad
Y_\Theta := \sqrt{\frac{8N_C}{3}}
\langle \Theta^+| \sum_{\alpha,\beta=4}^7d_{3\alpha\beta}
D_{+\alpha}D_{8\beta}|n\rangle 
\end{equation}
approach unity as $N_C\to\infty$ in the flavor symmetric case. 
In general they are computed from the eigenstates of the 
collective coordinate Hamiltonian, eq.~(\ref{evprob}).
The resulting width is shown for $N_C=3$ in fig.~\ref{fig_4} for 
the flavor symmetric case and the physical kaon--pion
mass difference.
\begin{figure}
\parbox[l]{10.0cm}{
\centerline{
\epsfig{file=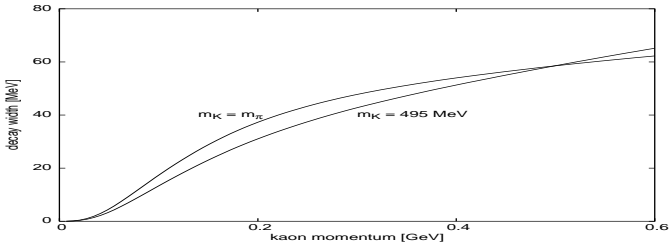,width=3.2cm,height=9cm,angle=270}}}~~~
\parbox[r]{4cm}{\vskip-0.5cm
\caption{\label{fig_4}\sl\footnotesize
Model prediction for the decay width, $\Gamma(\omega)$
of $\Theta^+$ for $N_C=3$ as function of the kaon momentum
$k=\sqrt{\omega^2-m_K^2}$, {\it cf.\@} eq.~(\ref{widthsb}). 
Top: $m_K=m_\pi$, bottom: $m_K\ne m_\pi$.
}}
\end{figure}
As function of momentum, there are only minor differences between
these two cases. Assuming the observed resonance to be the 
(disputed~\cite{Hi04}) $\Theta^+(1540)$ a width of roughly 
$40{\rm MeV}$ is read off from fig.~\ref{fig_4}~\cite{Wa05}.
It should be kept in mind that the general results on the treatment
of strange degrees of freedom are model independent but the numerical
results for the masses and the widths of pentaquarks are not.

\bigskip
\centerline{\large \bf Conclusion}
\smallskip

In this talk I have presented a thorough comparison~\cite{Wa05} between 
the bound state (BSA) and rigid rotator approaches (RRA) to chiral 
soliton models in flavor $SU(3)$. For definiteness I have only 
considered the simplest version of the Skyrme model augmented by 
the Wess--Zumino and symmetry breaking terms. However, this analysis 
merely concerns the treatment of kaon degrees of freedom. Therefore the 
qualitative results are valid for \emph{any} chiral soliton model.

A sensible comparison with
the BSA requires the consideration of harmonic oscillations in the
RRA as well. They can indeed be incorporated via the rotation--vibration 
approach (RVA), however constraints must be implemented to ensure that 
the introduction of such fluctuations does not double--count any 
degrees of freedom. The RVA clearly shows that the prediction
of pentaquarks is not an artifact of the RRA, pentaquarks are 
genu\-ine within chiral soliton models. Only within the RVA 
chiral soliton models generate interactions
for hadronic decays. Technically the derivation of this Hamiltonian
is quite involved, however, the result is as simple as convincing:
In the limit $N_C\to\infty$, in which the BSA is undoubtedly correct,
the RVA and BSA yield identical results for the baryon 
spectrum and the kaon--nucleon $S$-matrix. This identity
also holds when flavor symmetry breaking is included. This 
is very encouraging as it demonstrates that collective coordinate 
quantization may be successfully applied regardless of whether or not
the respective modes are zero--modes. Though the large $N_C$ limit
is helpful for testing the results of the RVA, taking only leading 
terms in the respective matrix elements is not trustworthy.

In the flavor symmetric case the interaction Hamiltonian contains 
only a single structure ($X_\Theta$) of $SU(3)$ matrix elements for 
the $\Theta^+\to KN$ transition. Any additional $SU(3)$ structure 
only enters via flavor symmetry breaking. This proves earlier 
approaches~\cite{Di97,El04} incorrect that adopted any possible 
structure that would contribute in the large $N_C$ limit and fitted 
coefficients from a variety of hadronic 
decays under the assumption of $SU(3)$ relations. That treatment 
yielded a potentially small $\Theta^+$ width from cancellations 
between different such structures even in the flavor symmetric 
case. The study presented in this talk thus clearly shows that 
it is not worthwhile to bother about the obvious arithmetic error 
in ref.~\cite{Di97} that was discovered earlier~\cite{We98,Ja04}
because the conceptual deficiencies in such width calculations are 
more severe. Assuming $SU(3)$ relations
among hadronic decays is not a valid procedure in chiral soliton
models. The embedding of the classical soliton breaks $SU(3)$ and 
thus yields different structures for different hadronic transitions. 
Especially strangeness conserving and changing processes 
are not related to each other in chiral soliton model treatments.

Even in case pentaquarks turn out not to be what recent
experiments have indicated, they have definitely been very beneficial
in combining the bound state and rigid rotator approaches and
solving the Yukawa problem in the kaon sector; both
long standing puzzles in chiral soliton models.

\bigskip
\centerline{\large \bf Acknowledgments}
\smallskip
I am grateful to the organizers for this pleasant workshop. I am 
very appreciative to Hans Walliser for the fruitful collaboration without 
which this presentation would not have been possible.

\end{document}